\documentclass[dvipdfmx]{revtex4}
\usepackage[dvipdfmx]{graphicx}
\usepackage{color}
\usepackage{ulem}
\newcommand{\bm}[1]{{\mbox{\boldmath $#1$}}}
\newcommand\erase{\bgroup\markoverwith{\textcolor{red}{\rule[.5ex]{2pt}{0.4pt}}}\ULon}
\begin{document}
\title{The training response law explains how deep neural networks learn}
\author{Kenichi Nakazato}
\affiliation{Bosch Center for Artificial Intelligence, Robert
  Bosch Japan, Tokyo, Japan\\
  Kenichi.Nakazato@jp.bosch.com\\
  Present Address: kenichi\_nakazato@sensetime.jp}
\date{\today}

\begin{abstract}
Deep neural network is the widely applied technology in this decade.
In spite of the fruitful applications, the mechanism behind that is
still to be elucidated. We study the learning process with a very
simple supervised learning encoding problem. As a result, we found a
simple law, in the training response, which describes neural tangent
kernel. The response consists of a power law
like decay multiplied by a simple response kernel. We can construct a
simple mean-field dynamical model with the law, which explains how the
network learns. In the learning, the input space is split into sub-spaces
along competition between the kernels. With the iterated splits and
the aging, the network gets more complexity, but finally loses its
plasticity.
\end{abstract}

\maketitle
\section{Introduction}
Our intelligence or nervous system changes its own shape through our
experience. In other words, the nervous system has a dynamics driven
by external stimuli. In the field of artificial intelligence, we
implement such a dynamical system with artificial neural networks.  In
a few decades, we actually have an astonishing progress in the
performance of the deep neural networks
\citep{dnn0,dnn1,dnn2,dnn3,dnn4,dnn5}. In theory, the deep
neural networks can approximate any functions in some conditions
\citep{approx0,approx1,approx2}. If we can optimize the network appropriately, we get the
network which captures the relation not only in the training data set
but also in unknown ones as the result of generalization
\citep{gen0,ntk0,ntk1}. In
spite of many such applications, our understanding on the mechanism
behind the deep neural network is very limited. We study the mechanism
of the training with very simple networks here.

In the training, the network parameters are adjusted along the input
and output relation in the given data set. In the framework of
supervised learning, we usually have a fixed and finite data set, the
pairs of input and output, $\{(\bm{x},y)\}$. The network realizes a
function, $f(\bm{x},\bm{w})$, expected to predict the given output. In general, we optimize the
network parameters, $\bm{w}$, to minimize the loss function, $L(f(\bm{x},\bm{w}),y)$, which evaluates
the difference between the predicted value, $f(\bm{x},\bm{w})$, and the given
output, $y$.
This optimization is called as training. In the optimization, we
usually rely on the gradient-based method, where the gradient of the
loss against the network parameter, $\nabla_w L(f(\bm{x},\bm{w}),y)$, gives the training step.
Actually, we can write the step as,
  \begin{equation}
    \Delta\bm{w}=\eta\nabla_w L(f(\bm{x},\bm{w}),y),
  \end{equation}
  where the parameter, $\eta$, is a parameter for adjusting the size of each learning step. This is called as a learning rate.

In the optimization, non-gradient-based algorithms, like random search, are usually not adopted. It is known that such algorithms are effective to avoid local optima, but it is believed that we do not have serious problems on the point \citep{no_locopt}. Under such an assumption, we can say gradient-based method is effective enough.
We call this optimizing process as learning dynamics.

As the simplest training step, we can consider a pair of data set,
$(\bm{x}_o,y_o)$. With an optimizing step of the parameters, the
predicted value, $f(\bm{x}_o)$, is
shifted a little, $f(\bm{x}_o)+\Delta$. We should note this shift has the effect on any
predictions, $f(\bm{x})+\Delta(\bm{x}_o,\bm{x})$. We call the shift,
$\Delta(\bm{x}_o,\bm{x})$, as a learning response. As the
accumulation of the learning responses, the network realizes the
generalized solution, if it works well. We can consider on a famous
image classification problem, CIFAR10\citep{cifar10}.
As an example, the problem is convenient for the explanation.
  In the classification problem, we have dataset with the pairs of an image and a label.
  Each image includes an object from 10 classes, airplane, car, bird, cat, deer and so on.
  The label shows the class and the network should give the answer for an input image.
When we train a network
with an image, $\bm{x}_c$, of a car, the learning response,
$\Delta(\bm{x}_c,x)$, is expected to
have the similar effects on predictions for other images of a car as well.
On the contrary, prediction for images without a car should not have similar one.

  We know deep neural networks can nicely do that in some ways, but we do not know so much on the process.
  In a few years, we have intensive studies on the neural tangent kernel theory\citep{ntk0,ntk1,ntk2,ntk3}.
  The theory says we can solve the learning response in an ideal case, where we have infinite neural units in a layer\citep{ntk0,ntk1}.
  In general, deep neural networks consist of layers with the units and the number of layers and units are finite.
  The theory cannot fully explain the real networks, therefore, but we should note some points here.
  In the ideal case, the response can be solved and even is constant during the training.
  On the contrary, the response is not constant with the finite network and it depends on the initial condition as well\citep{ntk2, ntk3}.

  We focus on the finite case as well, but another ideal problem is considered here.
  Since the real world data, like CIFAR10, is too complex to analyze the data space, we start with a much simpler one.
  For easier analysis, random problems are convenient, as other problems, e.g. TSP and k-SAT\citep{sopt,ccom}.
  We know the structure of the data space and have some techniques, like mean field approximation, in the random cases.
  By considering the ensemble of such random problems, we expect to get the fundamental understanding independent of a specific dataset instance.
  As a random problem, we consider random bit encoders, RBE.
  The input is 1d bit-string and the output is a binary value, 0 or 1.
  We can generate random problems in the setting and study its statistical features.

  In this paper, we show the law of learning response in a simple form.
  In the form, we can confirm a kind of aging.
  The response gradually diminishes along the training and varies
  its specificity to the input, $\bm{x}_o$. 
  With the law, we can construct a simple learning dynamics model for the dynamical understanding.
  It tells us that the dynamics is not a straight forward relaxation to the optimal solution.
  As we can notice, the prediction often shows back and forth dynamics.
  Our model gives an explanation for such a complex one.
  As a typical scenario, we can describe the dynamics as the iterated
  splitting process of the input space, ${\cal X}$. The split regions
  are encoded differently into the output
  space, ${\cal Y}$.
  We also discuss on the impact of our findings on the machine learning in general.

\section{Model} 
We study the training response of the neural network, $f(\bm{x},\bm{w})$,
with the loss function, $L$. The network parameter, $\bm{w}$, should
be optimized through the training.
In the gradient-based optimization, we can write the update process as
the following,
\begin{equation}
f(\bm{x},\bm{w}+\bm{\delta})=f(\bm{x},\bm{w})+\nabla_wf(\bm{x},\bm{w})\cdot\bm{\delta}.
\end{equation}
As the training step, $\bm{\delta}$, we can use of the gradient,
\begin{equation}
\bm{\delta}=-\sum_i\eta\nabla_wf(\bm{x}_i,\bm{w})\frac{dL}{df(\bm{x}_i,\bm{w})},
\end{equation}
where the parameter, $\eta$, is the learning rate.
The network is usually trained with training dataset, $\{\bm{x}_i\}$, at each step.
However, we focus on a training response, $\Delta(\bm{x}_o,\bm{x})$, for simplicity.
Once we understand its behaviour, we should understand the whole training dynamics as the sum of them.

The shift of the prediction, training response, can be described like,
\begin{equation}
\Delta(\bm{x}_o,\bm{x})\propto-\nabla_wf(\bm{x},\bm{w})\cdot\nabla_wf(\bm{x}_o,\bm{w})\frac{dL}{df(\bm{x}_o,\bm{w})}=-\Theta(\bm{x}_o,\bm{x})\frac{dL}{df(\bm{x}_o,\bm{w})}.
\end{equation}
The term, $\Theta(\bm{x}_o,\bm{x})$, is called as neural tangent
kernel, NTK, and it can be asymptotically solvable if the layer of the
neural network has infinite neurons in each layer\citep{ntk0,ntk1}.

If we can assume the loss function is just the mean squared error, the
training response can be written like,
\begin{equation}
\frac{\Delta(\bm{x}_o,\bm{x})}{y_o-f(\bm{x}_o,\bm{w})}\propto\Theta(\bm{x}_o,\bm{x}).
\end{equation}

To study the structure of the training response or NTK, we use a very
simple neural network with $N$ layers,
\begin{equation}
  y=f(\bm{x})\equiv f_N\circ\cdots\circ f_0(\bm{x}),
\end{equation}
where each layer is described as a function, $f_i$.
In general, we have $M_{in}$ input channels and $M_{out}$ output ones in each layer.
We can write the layer function with the input, $a_j$, and the output, $a_i^*$,
\begin{equation}
  a_{i}^*=g(\sum_j f_{ij}(a_{j})),
\end{equation}
where the function, $g$, is a non-linear one and we call it as activation function.
As the function, here, we use ELU or Relu\citep{elu0}.
As for the function, $f_{ij}$, we can use a linear transformation, $\bm{w}\cdot\bm{a}+b$, or a linear convolution.
To reduce the number of parameters, we usually use the convolution and it is called
as convolutional neural network\citep{cnn0,cnn1,cnn2}.
Here, we use a simple network with $N$ convolution layers and fix the number of channels, $M$, for simplicity.
Finally, we should output prediction, $y$, and use a fully connected linear transformation with the sigmoidal activation as the last layer.

  As the training dataset, we use a pair of 1D bit-string and a binary output.
  In other words, we want the network to learn the encoding, relation between the bit-string and the output.
  As a random problem, we can randomly sample a bit-string with the fixed-size, $s=16$.
  The output is usually a deterministic value in practical applications, but
  we assumue an encoding distribution, $p(y)$, and sample a binary value at each training step.
  This means the distribution can be written like, $p(y)=p_0\delta_u(0)+p_1\delta_y(1)$.
  To be noted, we can interprete this type of encoding as noised training.
  If either of one, $p_0$ or $p_1$, is 0 or 1, it means pure signal without any noise.
  We call this type of encoding problem as random bit-string encoder, RBE, here.

We firstly study the structure of the training response with the
case of single training data point and show a simple form which
describes the response.
Then the interaction between the training responses is studied with
the case of two points.

The simple form of training response tells us that the learning
dynamics can be described with very simple mean-field equations in our
case, RBE. We also demonstrate the learning dynamics and compare with
the simple dynamical model.

\section{Results}
\begin{figure}
\includegraphics[width=0.8\textwidth]{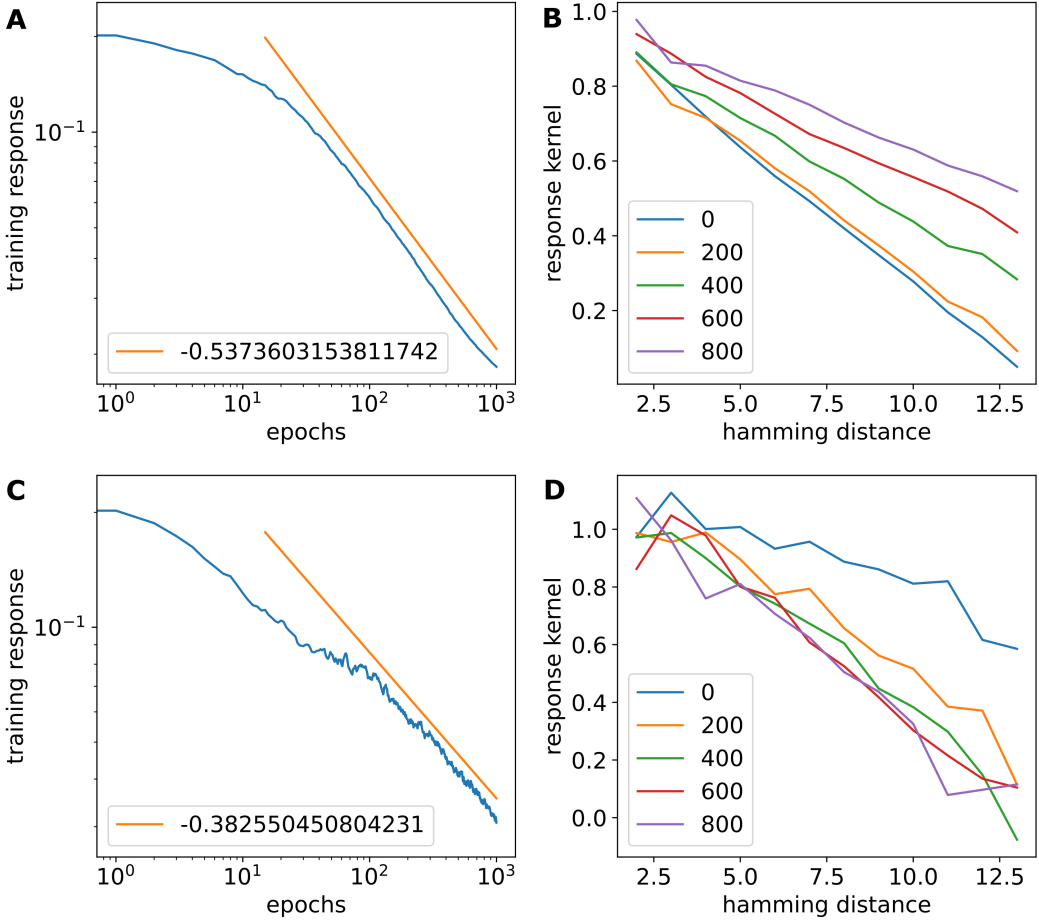}
\caption{Training response. Normalized training response consists of
the decay in the size and the kernel. (A) The decay for the case of
training without any bias. (B) The training kernel for the case
without any bias. (C) The decay for the case of biased training. (D)
The training kernel for the case of biased training. The kernels are
shown along the training epochs, $e=0, 200, 400, 600, 800$. All results
are averaged over 100 times tests.}\label{F1}
\end{figure}

In Fig. \ref{F1}, we show the averaged response, $\Delta(\bm{x}_o,\bm{x}_h)$,
along the Hamming distance from the input, $\bm{x}_o$.
The Hamming distance, $H_{ij}$, can be described as, $H_{ij}=\sum|\bm{x}_i-\bm{x}_j|$.
  In other words, it means the sum of different bits between the 2 bit-strings.

The network consists of $N$ layers and each layer has $M$ convolutions filters.
Here we set the parameters, $N=3$ and $M=8$. The learning
 rate, $\lambda$, is $0.1$. We used SGD as the optimizer and implemented
 it with keras \citep{sgd0,sgd1,sgd2,sgd3,sgd4,sgd5}. We call this as the standard setting in this paper.
The averaged response can be expressed with a simple linear kernel and its size.
After the initial training phase, the size shows the power law like
decay. We can summarize this response in the following,
\begin{equation}
\frac{\Delta(\bm{x}_o,\bm{x}_h)}{y_o-f(\bm{x}_o)}\propto
e^{-\tau}K(\bm{x}_o,\bm{x}_h,e),
\end{equation}
where K is the response kernel. The dynamics is driven along the epoch, $e$,
  and the power law term, $e^{-\tau}$, means the decay along the training epochs.
The kernel is almost
linear against the Hamming distance between the training input,
$\bm{x}_o$, and the affected one, $\bm{x}_h$. The slope varies along
the training epoch, $e$, depending on the training distribution.
If the training is non-biased, $p(0)=p(1)$, the kernel slope is
gradually diminished along the training. On the contrary, the slope is
enhanced in the case of biased training, $p(0)/p(1)=1/4$.

\begin{figure}
\includegraphics[width=\textwidth]{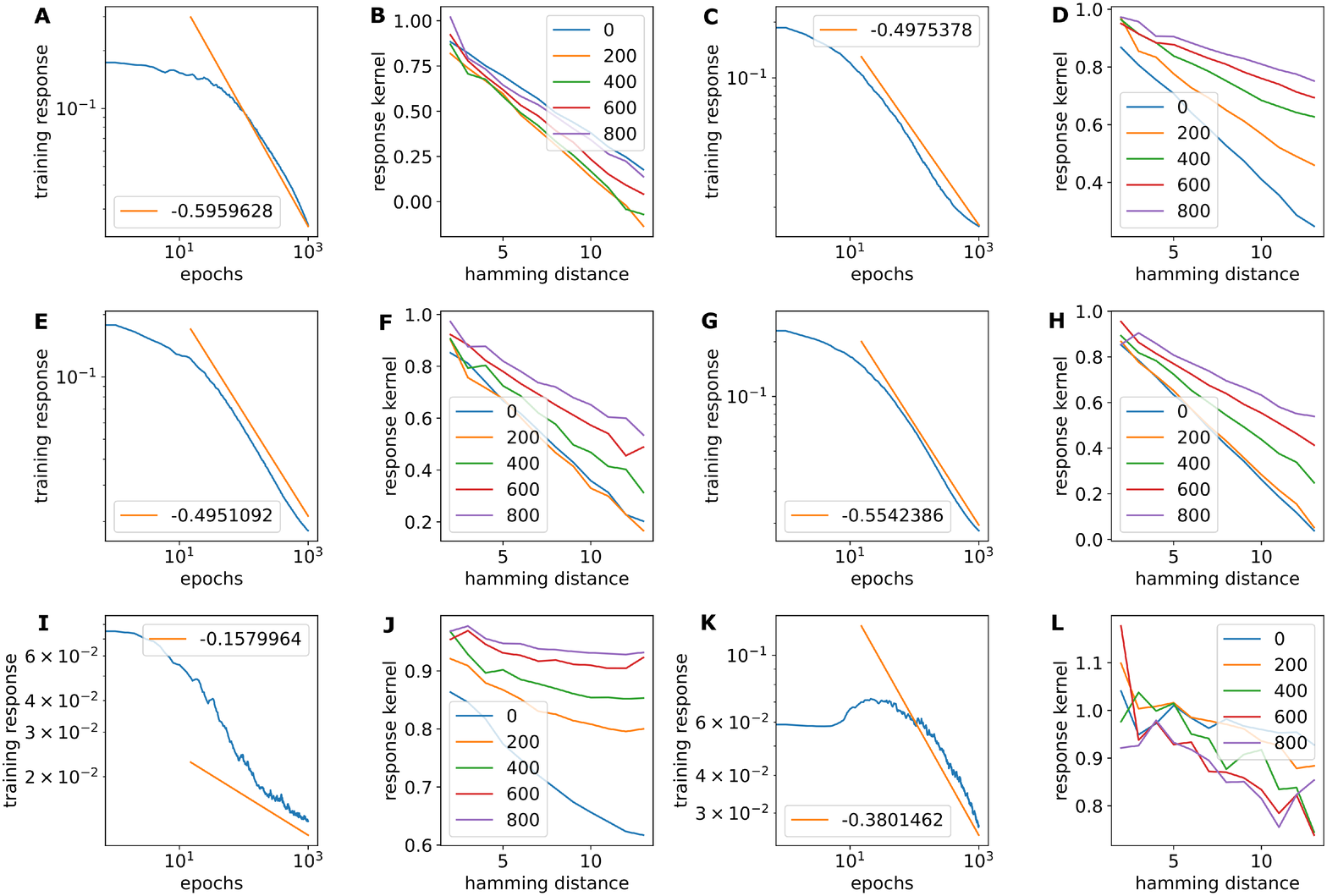}
\caption{Training responses for various conditions. In the upper two
lows, we show the cases with non-biased training for some parameter
settings. In (A),(B),(C) and (D), we show the case with the layer
number, $L=1$ and $5$. In the second low, (E),(F),(G) and (H), the
results with the number of convolution filters, $M=4$ and $16$, are
shown. In the bottom, we show the results with the other activation
function, Relu. The results with non-biased training are shown in (I)
and (J). The cases with biased training are in the right, (K) and (L).}\label{S1}
\end{figure}

We also tested the various conditions, shown in Fig. \ref{S1},
and the results can be expressed with the same law. We tested the
numbers of layers, $L=1$ and $5$, and the number of convolution
filters, $M=4$ and $16$, with the standard setting.
In addition, we tested one more activation function, Relu \citep{relu0}.
For all of the cases, the results with non-biased and biased training
are shown.
We can confirm the same tendency in those results and the simple
expression can be applied for the cases as well. The aging, decay in
the size of the response, shows clearer power law like behavior with
ELU rather than Relu.

\begin{figure}
\includegraphics[width=0.8\textwidth]{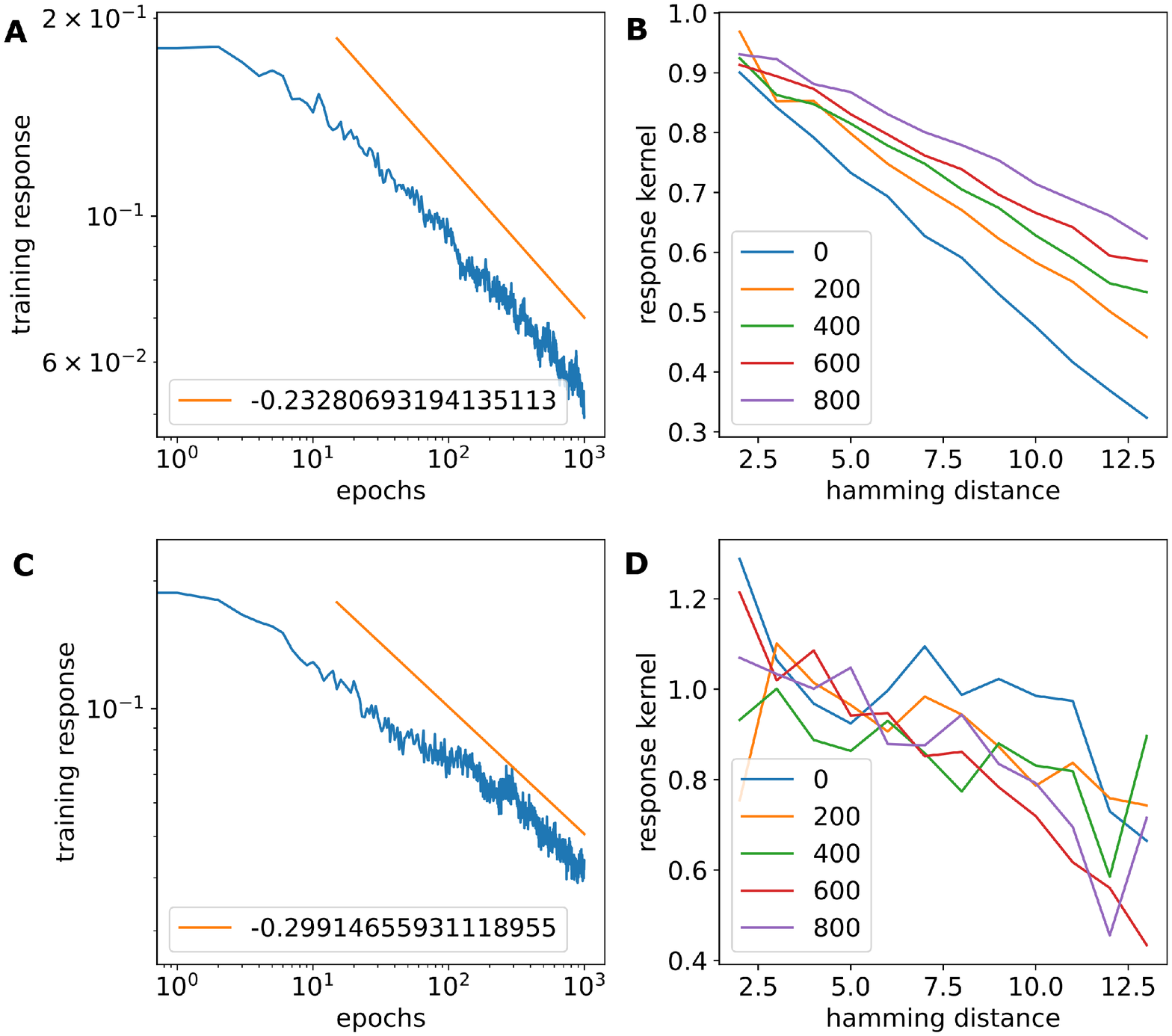}
\caption{The dropout kernel. The training response is shown with the
dropout effect for each convolutional layer. The dropout rate is
$0.2$. The network is the standard one. The training is non-biased, in
(A) and (B), and biased, in (C) and (D).}\label{S3}
\end{figure}

As a more complicated response, we show the results with the dropout
effect for each convolutional layer, in Fig. \ref{S3}\citep{do0,do1,do2}. The network is
the standard one, but the outputs from all of the mid-layers are
dropout with the rate, $0.2$. We should note the slopes of the
response kernels are smaller than the cases without dropout. Since the
dropout has the effect of coarse graining, the specificity of the
response kernels should be reduced and the results are reasonable.

\begin{figure}
\includegraphics[width=0.8\textwidth]{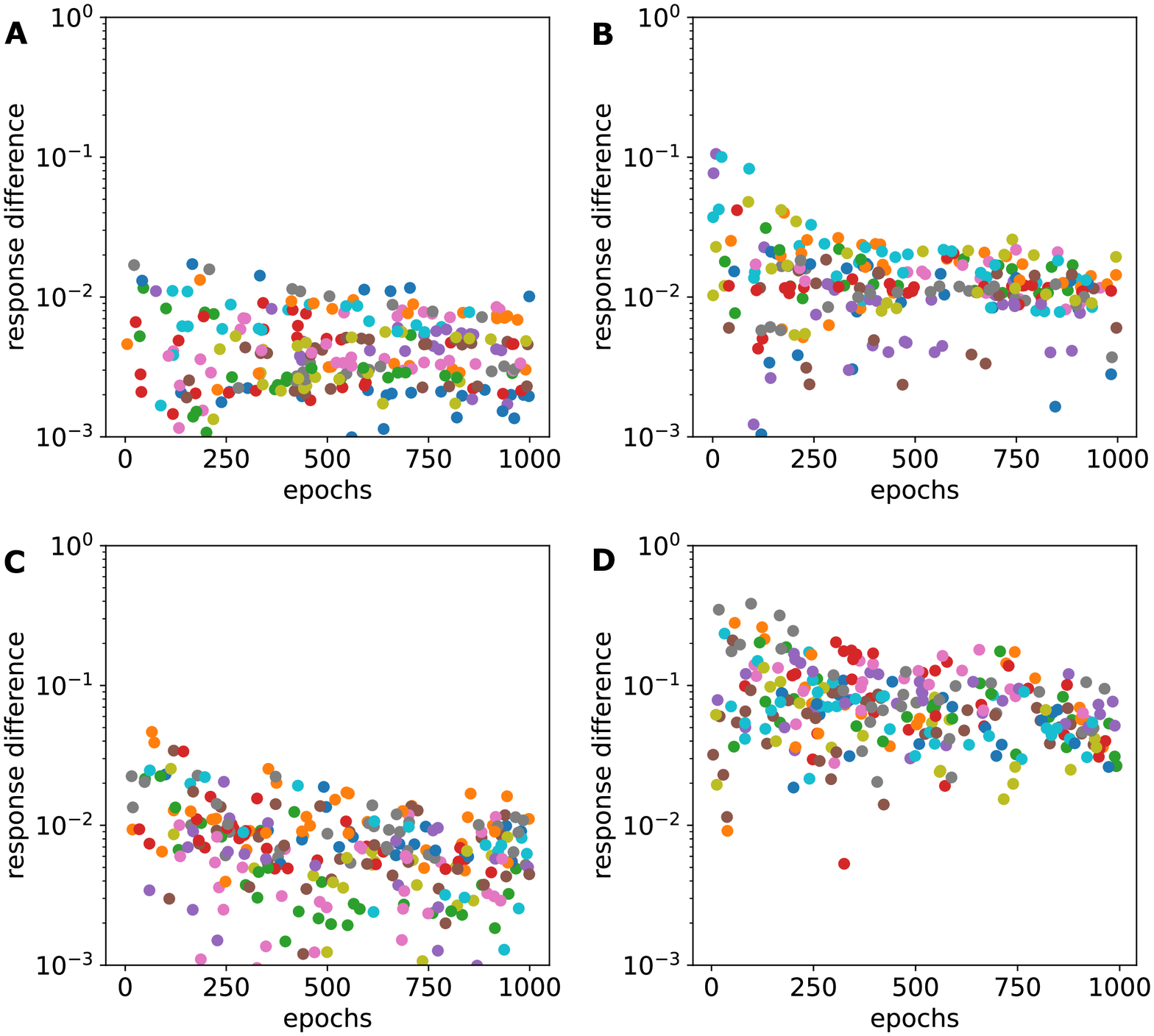}
\caption{The difference between the training responses for multiple
inputs. The size difference between the normalized training responses,
$\Delta(\bm{x}_1,\bm{x})$ and $\Delta(\bm{x_2},\bm{x})$, are shown as
scatter plots for 30 times tests. The mixing parameter is very small,
$\alpha=0.01$, for considering the case with almost one
way interaction. The Hamming distance between the two points is 1, in
(A) and (C) and 12, in (B) and (D). The training is
non-biased, in (A) and (B), and is biased, in (C) and (D).}\label{F2}
\end{figure}

We show the case with multiple training data points for understanding
the interaction of the responses, in Fig. \ref{F2}. As the simplest case, we
consider RBE with the two points training distribution,
$p(x,y)\propto((1-\alpha)\delta(\bm{x}_1)+\alpha\delta(\bm{x}_2))p(y)$.
As the encoding, $p(y)$, we use the non-biased distribution,
$p(0)=p(1)$, and biased one, $p(0)/p(1)=1/4$.  We can
confirm different aging along the Hamming distances between the
inputs,  $\bm{x}_1$ and $\bm{x}_2$. This means a training effect for
the input, $\bm{x}_1$, is not restricted to the training response,
$\Delta(\bm{x}_1,\bm{x})$, but also to others, like
$\Delta(\bm{x}_2,\bm{x})$, depending on the distance between the inputs.
If the distance between the points is small, the agings are enhanced
with each other, but the enhancement is very limited in the distant case.
This shows the training with an input, $\bm{x}_1$, has a strong effect on
the points around that, not only in the prediction, $f(\bm{x}_2)$, but also in the
training response, $\Delta(\bm{x}_2,\bm{x})$, as well.
On the other hand, the interaction effect is very limited if the points are
distant enough.

As the demonstration of the training response law, we consider another
simple situation, 
\begin{equation}
p(x,y)\propto\sum_{g_0}\delta_x(\bm{x}_i)\delta_y(0)+\sum_{g_1}\delta_x(\bm{x}_i)\delta_y(1).
\end{equation}
We have two groups of training dataset, $g_0$ and $g_1$.
In the group, $g_0$, $n$ training inputs, $\bm{x}_i$, are encoded into $0$ and $m$
inputs, $\bm{x}_j$, are encoded into $1$ in the group, $g_1$.

The short term learning dynamics of the prediction can be written in
the following,
\begin{equation}
\frac{df_i}{de}\propto\sum_{g_0}(0-f_j)\Delta(\bm{x}_j,\bm{x}_i)+\sum_{g_1}(1-f_j)\Delta(\bm{x}_j,\bm{x}_i).
\end{equation}
We can simplify it with the mean-field approximation under the sparse
condition,
\begin{equation}
\frac{df_i}{de}\propto -f_0\rho_0+(1-f_1)\rho_1,
\end{equation}
where the mean-field densities, $\rho_0$ and $\rho_1$, and averaged
predictions, $f_0$ and $f_1$, are used. The responses would have
non-linear interaction with each other in the dense condition, but we can treat
it as just the linear summation in this case. The aging effect is ignored in
the case of short-term mean-field dynamics, but the equation is not so different
except for the time and spatial scale of the dynamics even in the case
requiring the original equation.
The mean-field model means the dynamics is determined
just along the ratio between the densities, $\rho_0$ and $\rho_1$. The loss of
majority decreases in the first and the other one decreases later.

\begin{figure}
\includegraphics[width=0.8\textwidth]{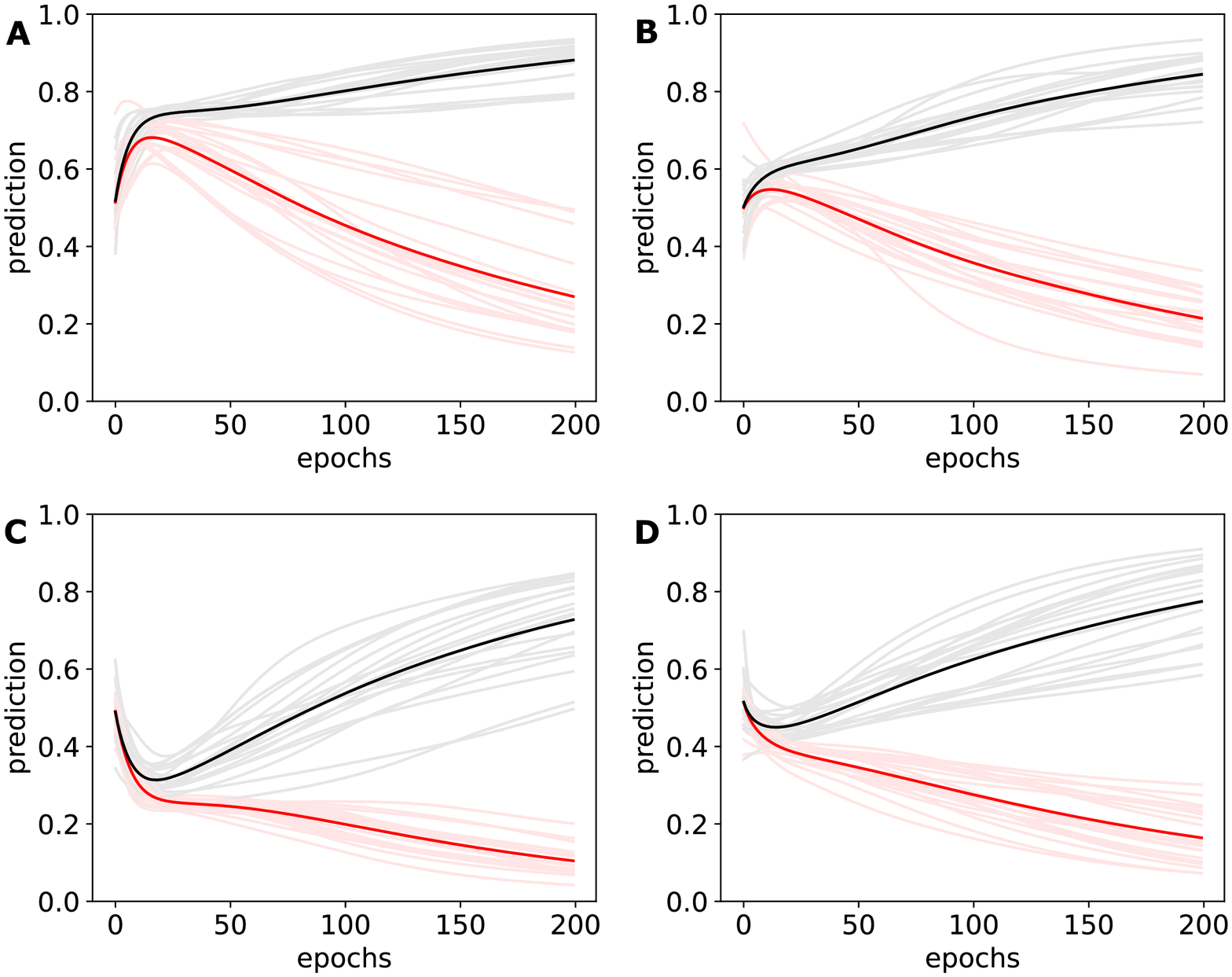}
\caption{Learning dynamics of RBE. The prediction curves are shown
with the training distributions, $\rho_0/\rho_1=1/3$, $2/3$, $3$ and
$3/2$, in order, (A), (B), (C)
and (D). The black ones show the dynamics of 1 encoded group and red
ones are 0 encoded. The bold curves show the averaged one of those 20
times tests.}\label{F3}
\end{figure}

We show the corresponding cases with the neural network, in Fig. \ref{F3}.
The network has the standard setting. The training data set consists
of 20 pairs, $\{(\bm{x}_i,y_i)\}$ $(i=0,\cdots,19)$. As the inputs,
random bit-strings are generated and those are assigned into $g_0$ or
$g_1$ with the given ratio, $\rho_0/\rho_1$. In the plot, we show the
time series of each prediction, $f_0$ and $f_1$, with the light color.
The bold ones show the average of those dynamics. As we can confirm,
all of the lines can be divided into two phases. In the first, all of
them go to to the majority side and then the minorities turn its
direction into the other side. In other words, the dynamics is
determined just along the ratio, $\rho_0/\rho_1$.

\begin{figure}
\includegraphics[width=0.8\textwidth]{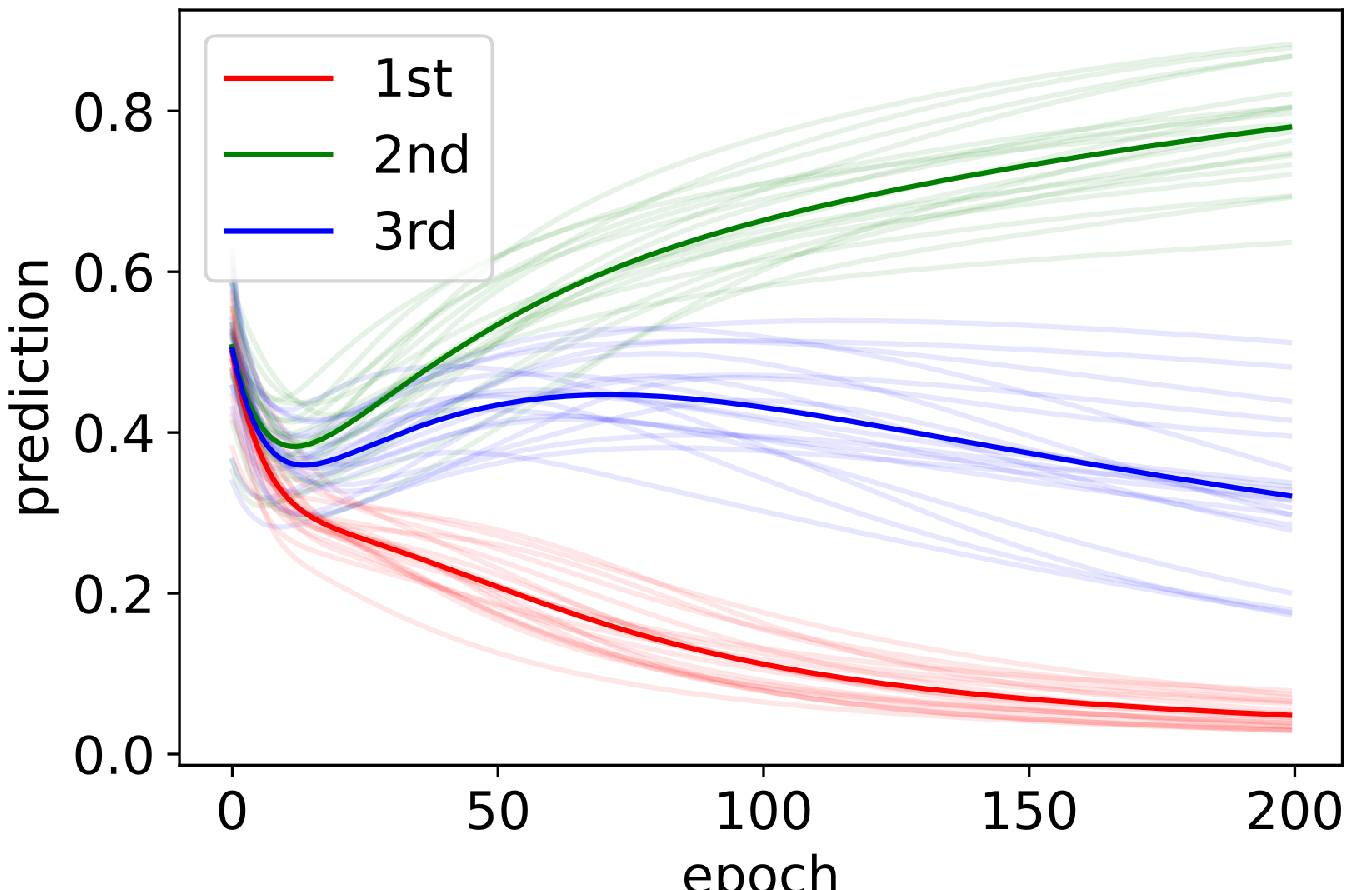}
\caption{Learning dynamics of RBE with a complicated training
distribution. The first group, $g_0$, is colored in red. The second
and third, $g_1$ and $g_2$, are colored in green and blue. The groups,
$g_1$ and $g_2$, are localized within the Hamming distance, $H_1=6$.
The group, $g_2$, is localized within the distance, $H_2=4$.}\label{S2}
\end{figure}

We also show hierarchically distributed case, in Fig. \ref{S2}, as an example
of a more complicated one.
The training data set consists of three groups, $g_0$, $g_1$ and $g_2$, and those are
hierarchically distributed. All of the input bit-strings have the length,
$s=16$, and those are randomly distributed, but the second and third
groups, $g_1$ and $g_2$, are localized,
\begin{equation}
H(\bm{x}_i,\bm{x}_j)\leq H_1\ \ (\forall i,j\in g_1\cup g_2),
\end{equation}
where the Hamming distance, $H(\bm{x}_i,\bm{x}_j)$, is used.
Furthermore, the bit-strings the group, $g_2$, are localized in a
narrower area,
\begin{equation}
H(\bm{x}_i,\bm{x}_j)\leq H_2\ \ (\forall i,j\in g_2).
\end{equation}
In the fig. \ref{S2}, the restricted areas have the sizes, $H_1=6$ and
$H_2=4$.
Among those groups, two ones, $g_0$ and $g_2$, are encoded as 0 and
the other one, $g_1$ is encoded as 1.

In the figure, we show all of the averaged predictions, $f_i$, in the light
color and bold ones show the average of them within the each group. We
iterated this for 20 times and the results are shown.
Since the encoding majority is 0, all curves show a decrease at the
first. In the second, the minor encoding group, $g_1$, turns its
direction, but the last group, $g_2$, is accompanied with the
movement. This is because the group $g_1$ is the majority in the
localized groups. Finally, after some epochs, the last one, $g_2$,
gets released from the group, $g_1$, because the loss of the group,
$g_1$, is reduced enough and the training response,
$\Delta(\bm{x}_1,\bm{x}_2)$, is diminished as well.
In sum, the input space is classified as the same one at the first
and the space is split into two classes in the second. The newly split class is
split again along its encoding finally.

\section{Discussion}
In supervised learning, we train a model with the given data. The
model is optimized to satisfy the input-output relations in the data
and it is generalized in someway. To understand the process, we
studied the response function, $\Delta(\bm{x}_0,\bm{x})$, and studied
its structure with a very
simple problem, RBE.
  As a model network, we studied a very simple convolutional neural network
  and found a training response law. That consists
of a power law like decay and the response kernel. This gives NTK a
simple form. It tells us the training dynamics with the finite sized network.


Since our model network is very simple, the response kernel is very simple linear one.
However, the kernel can have a more complicated non-linear form, if
the network architecture is more sophisticated. In reality, we can
show the kernel for the network with drop-out layers, in fig. \ref{S3}.
It shows broader peak around the input, $\bm{x}_o$.  We believe
the response kernel analysis is effective for more complicated
networks, in the similar way.

We know NTK can be analytically solved with the central limit theorem,
if each layer has infinite units \citep{ntk1}. It has no dependence on its
dynamic state. On the contrary, we can confirm aging in its dynamics,
but the form is still kept simple, in our cases. Since our network
is finite one, the parameter fluctuates toward the area with less training
response, driven by kinesis \citep{kinesis}. The shape of response kernel is very
similar with NTK of infinite linear network, but it shows some
dependence on the dynamics. In the short term, we can derive much
simpler dynamical model with mean-field approximation. We can easily
understand how the network learns with our simple situation, RBE.
Initially the whole input space is encoded into the majority. After
the initial encoding, minority clusters turn over its encoding. In
other words, the input space is roughly split into the majority and
minorities. This type of space split can happen in an iterated manner.
Finally, we get to an optimal split, if the network has enough degree
of freedom.

Our view tells us how deep neural networks can get to generalized and
non-overfit solutions. The network is optimized to minimize the
loss, which shows the error only for the training data set.
 In other words, we have no explicit loss for inputs in general.
  Rather, all we can know is the loss for training data set.
  We hope the generalization is achieved in a implicit manner.
  The training response shows the implicit effect.
  Since the training is usually done with a lot of training data points,
  the generalization can be seen as a result of the competition between the training responses of them.

As we noticed, the dynamics is similar with that of kernel machines
\citep{kernel0,kernel1,kernel2}. In it, we optimize the linear network,
\begin{equation}
y=g(\sum_j a_jK(\bm{x},\bm{x}_j)+b),
\end{equation}
where the activation function, $g$, outputs the prediction along the
network parameter, $a_j$ and $b$.
In this model, the vector of kernel similarity,
$Kij=K(\bm{x}_i,\bm{x}_j)$ or $\bm{K}_i=(K_{i0},\cdots,K_{in})$,
constitutes the training input. Since this network is very simple linear one,
  the response kernel would be a linear one, as we shown.
  In the kernel machine, we need sophisticated design of the kernel similarity, therefore.
  On the other hand, in the case of deep neural networks, we do not need such a sophisticated kernel design.
  Instead, the network architecture would realize it as a response kernel and the training dynamics.
  As we shown, the dynamics is not so complex, if we can ignore the aging or interaction of the kernels.
  However, it may be much more complex, if those effects cannot be ignored.

In our results, the response kernels are shown with non-deterministic
encoding, $p(0)>0$ and $p(1)>0$, but we already tested the case
with deterministic encoding, in Fig. \ref{A1}.
Especially, the aging of the response size shows a clear power law.
However, it was hard to confirm the convergence of the response kernels.
In this deterministic case, the aging exponent is larger than non-deterministic ones.
  The kernels seem to be divergent as the training proceeds.
  The diverged kernel would have no significant effects on the dynamics therefore.
  Rather, the dynamics would be determined with mainly non-diverged one.

  We studied a simple problem, RBE, but the same results can be observed in a real problem.
  We tested the training response with a known dataset, MNIST\cite{mnist}.
  We want to classify images of hand-written digit, $0, \cdots, 9$, along their digits.
  To focus on the training response against such real images, we trained the same type of convolutional network with a pair of randomly sampled MNIST image and encoding distribution, $p(y)$, in the similar way with the RBE.
  The results are shown in FIG. \ref{mnistRS} and \ref{mnistRK}.
  We can confirm both the power law decay and the distance dependent response kernel.
  The kernel is not so clear linear one but monotonically decreasing one.
  To be noted, we can observe some isolated responses at the same time.
  This suggests a chaotic phase depending on the initial condition\citep{ntk2,ntk3}.
  On the contrary, the aging seems to be more stable, but we need more tests to have a clear conclusion.

  In our problem setting, RBE, we originally intended to analyse the thermal equilibrium under a noised training.
  In other words, we expected to describe the training dynamics as a quasistatic process,
  which can be characterized with the noise ratio.
  However, we unexpectedly found the aging process instead of the equilibrium.
  As we discussed, the training dynamics should be directed to zero-response state against the training stimuli.
  That is because the residence time should be maximized under such a stochastic dynamics.
  We guess the network is redundant enough to realize not only the input output relation, but also such a no response state.

We studied the non-linear complex dynamics of the neural network with
the simple situation. As we know, such a complex system often has a very
simple description in spite of the complexity in its appearance
\citep{baryam,newman}. As one
more example of compex system, we found the simple law in the neural
network. This type of simplicity suggests the universality of the
mechanism in the variety of complex phenomena. We may find the same
type of algorithmic aging in the nervous system or complex dynamical
networks in general \citep{haken}. As other studies on complex systems, the
algorithmic matters should be fruitful subjects to promote
understanding of complexity in general \citep{sopt}.

\section*{Appendix}
\begin{figure}
\includegraphics[width=0.8\textwidth]{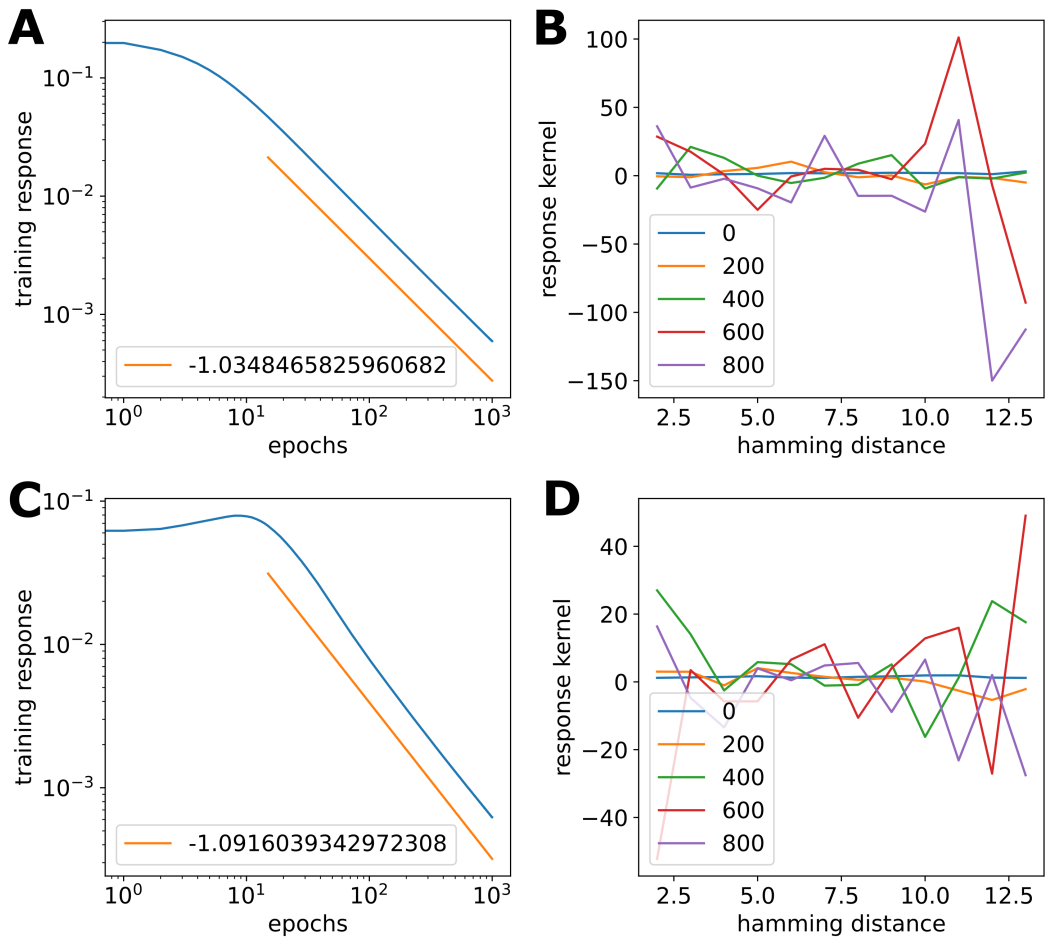}
\caption{Training responses for deterministic encoding. The aging of
the response size with the deterministic case, $p(0)=1$ and $p(1)=0$,
in (A). The response kernels are in (B). In (C) and (D), the training
response with the activation function Relu are shown. The network
setting is tha standard one. The kernels, in (B) and (D), are not
converged, but diverged along the training epochs.}\label{A1}
\end{figure}

The training responses for the deterministic encoding, $p(0)=1$ and
$p(1)=0$, are shown in the Fig. \ref{A1}. The network is the standard
one. We show the results for the activation functions, ELU and Relu.
In the size, we can confirm the power law, but the kernels do not
show convergence. Since the network has the sigmoid function as the activation
function for the output, convergence in the output does not mean that
of network parameters. This result suggests the fragility of the
response kernel and the prediction in the end.

\begin{figure}
\vspace{-5cm}\includegraphics[width=1.0\textwidth]{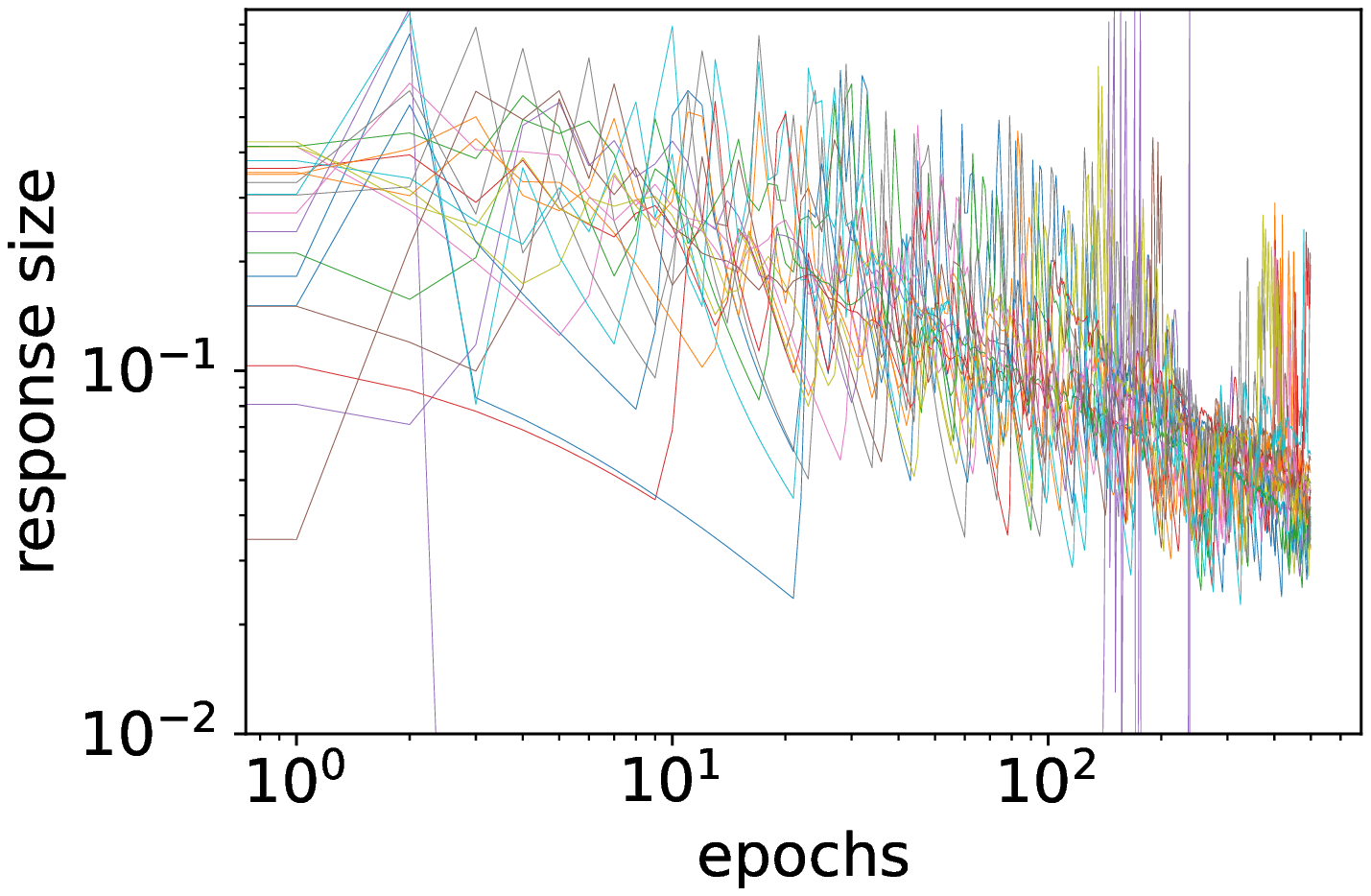}\vspace{-5cm}
\caption{The aging of the training response with MNIST.
  We tested the training 20 times and the sizes are plotted.
Each plot for a training sequence is colored differently.
}\label{mnistRS}
\end{figure}

\begin{figure}
\vspace{-5cm}\includegraphics[width=1.0\textwidth]{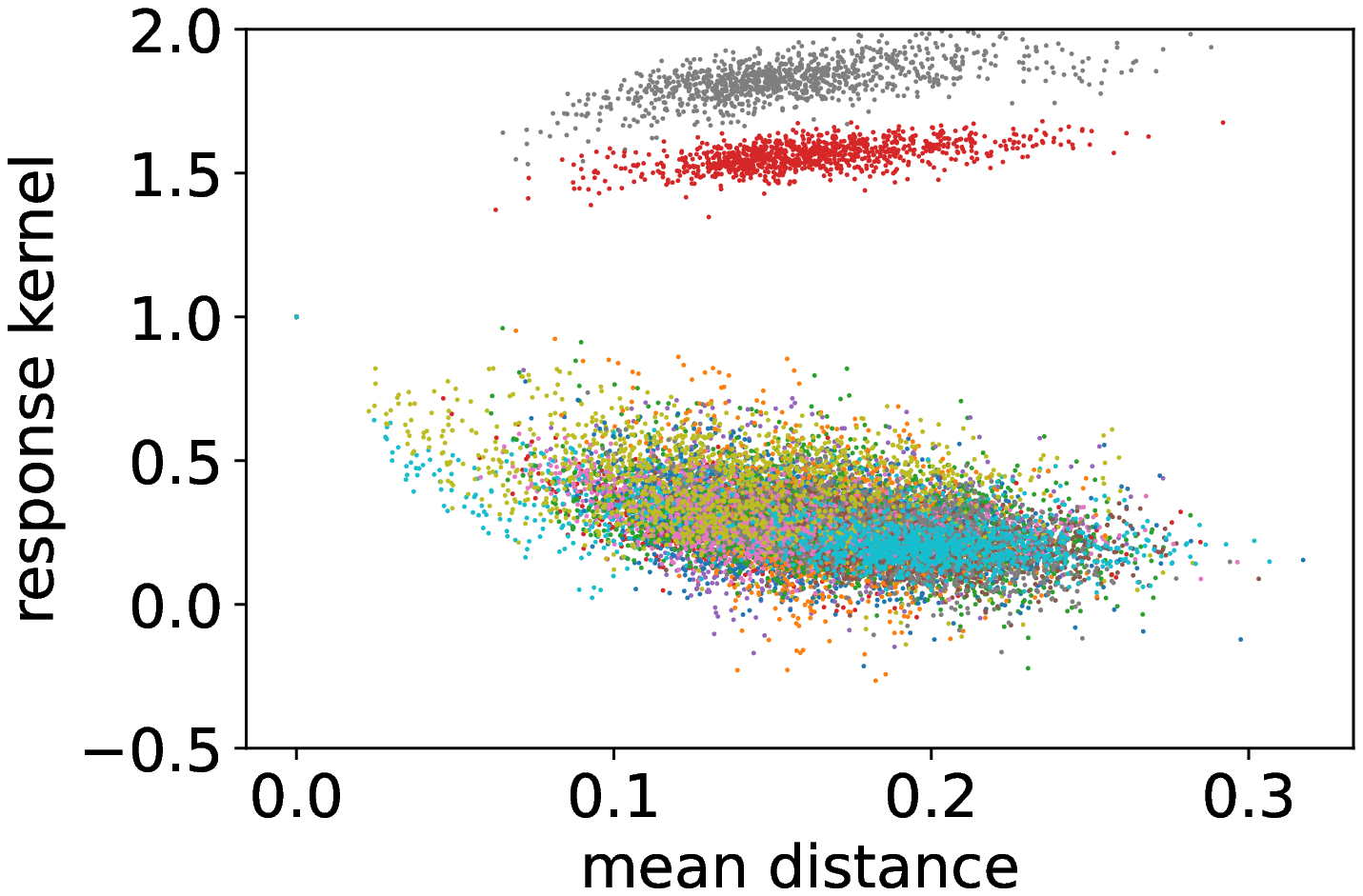}\vspace{-5cm}
\caption{The response kernels with MNIST.
  We tested the training 20 times and the response kernels are plotted.
We plotted 1000 samples after the training of 500 epochs.
The 20 trainings are colored differently.
}\label{mnistRK}
\end{figure}

As an realistic example, we show the training response of the same type of network, which is trained with
MNIST\citep{mnist}.
As the encoding distribution, we used the ratio, $p(0)/p(1)=1/5$.
In the FIG. \ref{mnistRS}, the aging is shown and we can observe a power law decay.
In this case, the network has 2 dimensional convolution along the input image size.
The activation function is ELU and the other network parameters are the same as the standard one.
In the FIG. \ref{mnistRK}, the response kernels are shown.
The horizontal axis is the mean distance between two images, $<|\bm{x}_i-\bm{x}_j|>$.
In other words, the plot shows the response kernel, $\Delta(\bm{x}_i,\bm{x}_j)$.
After the training of 500 epochs, we randomly sampled 1000 from 60000 images and plotted the response.

\begin{acknowledgements}
This work was originally started from the discussions with M. Takano and H.
Kohashiguchi and partially supported by PS/PJ-ETR-JP.
\end{acknowledgements}

\bibliography{aaa}
\end{document}